\documentclass[
aps,%
12pt,%
final,%
notitlepage,%
linenumbers,%
oneside,%
onecolumn,%
nobibnotes,%
nofootinbib,%
superscriptaddress,%
noshowpacs,%
centertags]%
{revtex4}

\usepackage{url}
\usepackage[english]{babel}

\usepackage[hidelinks]{hyperref}

\usepackage{graphicx}
\usepackage{amsmath}

\allowdisplaybreaks

\newcommand{\B}{\mathcal{B}}
\newcommand{\T}{\mathcal{T}}
\renewcommand{\Im}{\mathrm{Im}}
\newcommand{\eff}{\mathrm{eff}}
\newcommand{\GeV}{\text{GeV}}
\newcommand{\MeV}{\text{MeV}}

\newcommand{\PI}{\mathrm{PI}}
\newcommand{\WS}{\mathrm{WS}}
\newcommand{\ps}{\text{ps}}
\newcommand{\s}{\mathrm{s}}

\begin{document}

\author{\firstname{A.~V.}~\surname{Berezhnoy}}
\email{Alexander.Berezhnoy@cern.ch}
\affiliation{SINP of Moscow State University, Russia}

\author{\firstname{A.~K.}~\surname{Likhoded}}
\email{Anatolii.Likhoded@ihep.ru}
\affiliation{Institute for High Energy Physics, Protvino, Russia}
\affiliation{Moscow Institute of Physics and Technology}

\author{A.~V.~Luchinsky}
\email{alexey.luchinsky@ihep.ru}
\affiliation{Institute for High Energy Physics, Protvino, Russia}

\title{Doubly heavy baryons at LHC}
\begin{abstract}
  The theoretical analysis of  production, lifetime, and decays of doubly heavy baryons is presented. The lifetime of  $\Xi_{cc}^{++}$ baryon recently measured by  the LHCb Collaboration is used to estimate the lifetimes of other doubly heavy baryons. The production and the possibility of observation  of $\Xi_{bc}$ baryon at LHC are discussed.
\end{abstract}

\maketitle

\section{Introduction}
\label{sec:intro}

Doubly heavy baryons are extremely interesting objects that allow us to take a fresh look at the problems of the production and hadronization of heavy quarks.
These baryons consist of two heavy and one light quarks and therefore, unlike ordinary heavy baryons, are characterized by several scales at once:
\begin{equation}
m_{Q_{1,2}} \gg m_{Q_1} \cdot v, m_{Q_2}\cdot v \gg \Lambda_\mathrm{QCD},
\end{equation}   
where  $m_{Q_1},m_{Q_2}$ are masses of heavy quarks, and $v$ is there velocity inside the quarkonium.
For clarity, one can go to the coordinate representation and select a specific family of baryons. Thus, for a baryon $\Xi_{bc}$ containing $ b $ - and $ c $ -quarks simultaneously, the scales are ordered as follows:
\begin{equation}
\lambda_b : \lambda_c :r_{bc} : r_{QCD} \approx 1 : 3: 9 :27,
\end{equation}
where $\lambda_Q=1/m_Q$ is a Compton length of quark, 
$r_{bc}\sim 1/(v\cdot m_Q)$ is heavy quark size, 
$r_\mathrm{QCD}=\Lambda_\mathrm{QCD}$ is a scale of nonperturbative confinement~\cite{Kiselev:2001fw}.

It is worth to mention,  that a  baryon  with one heavy quark  is characterized by only two scales, namely, the mass of the heavy quark and $\Lambda_\mathrm{QCD}$.
In the limit $m_{Q_1}, m_{Q_2}  \rightarrow \infty$  a heavy diquark interacts with a light quark as heavy anti-quark and, therefore, it is quite natural to subdivide calculating the characteristics of doubly heavy quarkonium in two stages: the calculation of the properties of the heavy diquark and the subsequent calculation of the properties of the system of quark-diquark~\footnote{
An alternative approach based on the direct  solution of the three-body problem is presented in~\cite{Albertus:2006wb,Albertus:2006ya}}.

The problems of production and decays of such systems was of interest to researchers for many years. But the last year was special because it was marked by the discovery of the doubly charmed $\Xi_{cc}^{++}$ baryon in the decay mode $\Lambda_c^+ K^- \pi^+ \pi^+$~\cite{Aaij:2017ueg}. The LHCb Collaboration observes hundreds of such particles.  This discovery was confirmed by the observation of decay $\Xi_{cc}^{++}\rightarrow \Xi_{c}^{+}\pi^{+}$~\cite{Aaij:2018gfl}.   This circumstance greatly revived the research activities in this direction. In this article we discuss the perspectives of further research of doubly heavy baryon states: there decays, 
productions and possibility of observation of excited states.

The rest of the paper is organized as follows. In the next section production of doubly heavy baryons  is considered. Section \ref{sec:lifetimes} is devoted to theoretical calculation of the lifetimes of the considered particles. Observation probability of these baryons is discussed in section \ref{sec:observ-prob} and finally the Conclusion will be given.

\section{Doubly heavy baryon production}
\label{sec:produciton}

It is natural to use a two-step procedure to produce  a doubly heavy baryon. In the first calculation step a doubly heavy diquark is produced perturbatively in the hard  interaction.   In the second step a doubly heavy diquark is transformed to the baryon within the soft hadronization process. 

Our calculation of doubly heavy diquark production were done within the following	 approach:
\begin{enumerate}
\item 
the color singlet model for doubly heavy mesons and the color triplet model for doubly heavy baryons;
\item
the contribution from scattering of sea heavy quark  and gluon ($ Q_1 g\to Q_1 + Q_2 + \bar Q_2$) does not take into account to avoid double counting;~\footnote{Furthermore, accounting these process is questionable for LHCb kinematic region due to rather small transverse momenta of doubly heavy system which is comparable with a heavy quark mass.}
\item the contribution of color  sextet state to baryon production is neglected.
\end{enumerate}

Quarks in color antitriplet $\bar 3_c$  attract each other and their  interaction can be described by the wave function in the framework of potential model, as well as the quark-antiquark  interaction in quarkonium.  
By analogy with quarkonium one can write  for the production amplitude of doubly heavy diquark:
\begin{equation}
A^{SJj_z}=\int T^{Ss_z}_{Q_1\bar Q_1 Q_2 \bar Q_2}(p_i,k(\vec q))\cdot
\left (\Psi^{Ll_z}_{ [Q_1 Q_2]_{\bar 3_c}}(\vec q) \right )^* \cdot
C^{Jj_z}_{s_zl_z} \frac{d^3 \vec q}{{(2\pi)}^3},\nonumber
\end{equation}
where $T^{Ss_z}_{Q_1 \bar Q_1 Q_2 \bar Q_2}$ is an amplitude of the hard production of two heavy quark pairs;\\
$\Psi^{Ll_z}_{[Q_1 Q_2]_{\bar 3_c}}$ is the diquark wave function (color antitriplet);\\
$J$ and $j_z$ are the total angular momentum and its projection on $z$-axis in the $[Q_1 Q_2]_{\bar 3_c}$ rest frame;  \\
$L$ and $l_z$ are the orbital angular momentum of $bc$-diquark and its projection on $z$-axis;\\
$S$ and $s_z$ are $ Q_1 Q_2$-diquark spin and its projection;\\
$C^{Jj_z}_{s_zl_z}$ are Clebsh-Gordon coefficients;\\
$p_i$ are four momenta of  diquark, $\bar Q_1$ quark and  $\bar Q_2$ quark;\\
$\vec q$  is three momentum of $Q_1$-quark in the  $Q_1 Q_2$-diquark rest frame (in this frame $(0,\vec q ) = k(\vec q)$).

Under assumption of small dependence of $T^{Ss_z}_{b\bar b c \bar c}$ on $k(\vec q)$ 
\begin{equation}
A \sim 
\int d^3q\,\Psi^*({\vec q})\left\{ \bigl.T(p_i,{\vec q}) \bigr|_{\vec q=0}+
\bigl.{\vec q}\frac{\partial}{\partial {\vec q}} T(p_i,\vec q) \bigr|_{\vec q =0}  +  \dotsb \right\}\nonumber
\end{equation}
and, particularly, for the $S$-wave states
\begin{equation}
A \sim R_S(0) \cdot \bigl. T_{Q_1\bar Q_1 Q_2 \bar Q_2}(p_i) \bigr|_{\vec q=0}, \nonumber
\end{equation}
where $R_S(0)$ is a value of radial wave function at origin.

\begin{figure}
  \centering
  \includegraphics[width=0.7\textwidth]{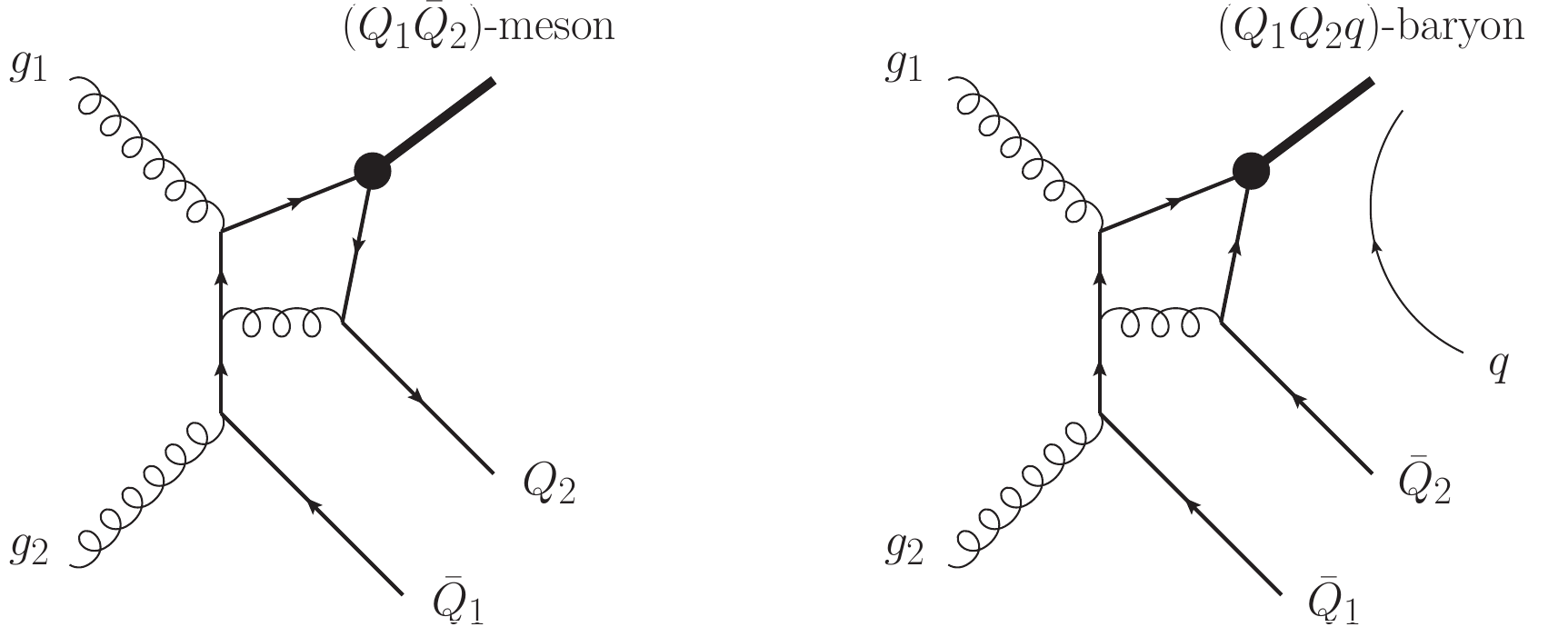}
  \caption{The example of analogous diagrams for $(Q_1 \bar Q_2)$-quarkonium production and for $(Q_1 Q_2 q)$-baryon production.}
  \label{fig:feynman_diagrams_example}
\end{figure}

In our early work~\cite{Berezhnoy:1998aa} we discussed the similarity of the production mechanisms of doubly charmed baryons and  the associative $J/\psi$ and the open charm in hadronic interactions. Indeed, both processes within  a single parton scattering approach are described by the similar sets of diagrams, because both ones  involve the production of four heavy quarks (see diagram examples in Fig.~\ref{fig:feynman_diagrams_example}). 
However, the  experimental data indicate the presence of contribution of double parton scattering (DPS), which dominates at LHC energies \cite{Aaij:2012dz}. Within the DPS mechanism two $c\bar c$ pairs are produced independently  in the different parton interactions. Such mechanism can contribute to the  associative $J/\psi + c$ production 
but one can hardly contribute to the process  $\Xi_{cc}$ production, because to produce doubly charmed baryon $c$ charm quark from different pairs are needed.\footnote{ However these is a research, where it was made an attempt  to expand the DPS model to the case of $\Xi_{cc}$ production~\cite{Kiselev:2016rqj} using quark-hadron duality approach.}  Thus  we currently tend to think,
that DPS mechanism contributes only to $J/\psi + c$ production.  This is why   the yield of  $\Xi_{cc}$ is essentially smaller, than the yield of the associative production of   $J/\psi$-meson and open  charm, whereas the yields of $B_c$ mesons and $\Xi_{bc}$ baryons should be comparable. 
Also it is worth to mention  that   $J/\psi + c$ cross section and $\Xi_{cc}$ cross section should have different dependence on the $pp$ interaction energy: DPS  cross section 
increases  faster than SPS. 

It should be noted that the doubly heavy diquark production
 can not be described within the fragmentation model due to the large contribution of non-fragmentation diagrams, which can not be interpreted as $b$-quark production followed by  the fusion of $b$-quark into $bc$-diquark. The same feature is inherent in the process of $B_c$-meson production.
This is not surprising because the production processes of $bc$-diquark production and $B_c$ production are described by the same set of the diagrams. The difference comes from  different color coefficients and different choice of values for $c$ and $b$ quark masses.  

The dominant contribution to the production cross under LHCb kinematics conditions comes from gluonic interaction,
as well as for the $B_c$ meson:
$$gg \to \Xi_{bc} +\bar b \bar c.$$

Our estimations for that process show that difference of yields of $\Xi_{bc}$ and $B_c$ is mostly determined by the 
difference  of wave functions:

\begin{equation}
\frac{\sigma_{\Xi_{bc}}}{\sigma_{B_c}}\sim \frac{|R_{[bc]_{\bar 3}}(0)|^2}{|R_{B_c}(0)|^2}
\label{eq:sigma_ratio}
\end{equation}

\begin{figure}
  \centering
  \includegraphics[width=0.4\textwidth]{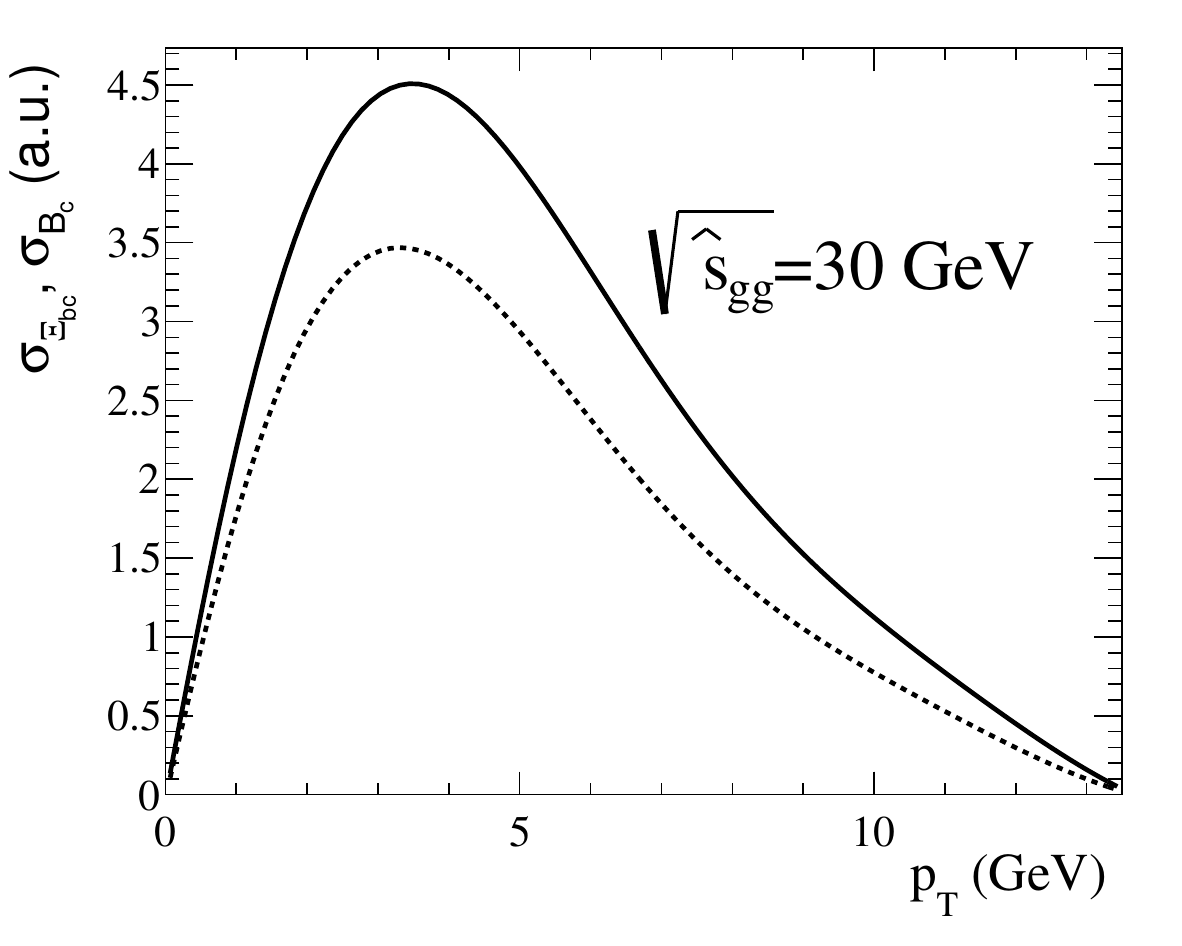}
  \includegraphics[width=0.4\textwidth]{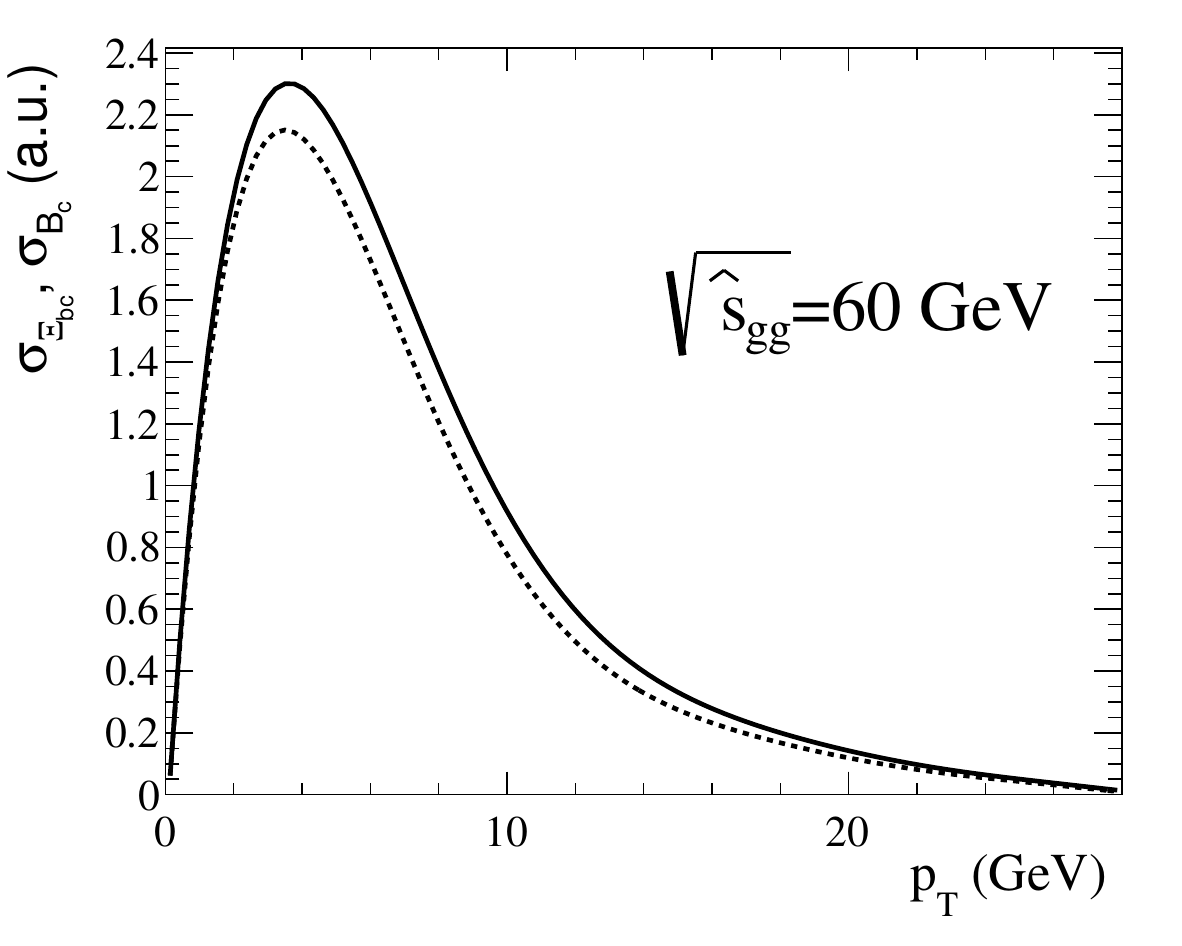}
  \caption{$\Xi_{bc}$ $p_T$ distribution 
 v.s. $B_c$ $p_T$ distribution for $\sqrt{s_{gg}}=30$ GeV and $\sqrt{s_{gg}}=60$ GeV, correspondingly. The same quark mass values are used for both estimations: $m_c=1.5 $ GeV and $m_b=4.8$ GeV. Also, for convenience of comparison, we put $|R_{B_c}(0)|^2$ and $|R_{[bc]_{\bar 3}}(0)|^2$ equal.}
  \label{fig:pt}
\end{figure}

Indeed, if one choose  the same quark mass values for the subprocesses $gg \to [bc]_{\bar 3} +\bar b \bar c$ and $gg \to B_c +\bar b \bar c$ and put $R_{[bc]_{\bar 3}}^2=R_{B_c}^2$ one can see that this process have very similar behavior on transverse momenta of doubly heavy system,   as it is shown in Fig.~\ref{fig:pt}, where we put $|R_{B_c}(0)|^2$ and $|R_{[bc]_{\bar 3}}(0)|^2$
 equal for convenience of comparison.

Of course, a color antitriplet of $bc$ system is not a $\Xi_{bc}$ yet. It should be somehow transformed to the  $bcq$ baryon.  The transverse momentum of light quark $q$  with mass $m_q$ is about
$\frac{m_q}{m_{\Xi_{bc}}}p_T^{\Xi_{bc}}$, where $p_T^{\Xi_{bc}}$ is a transverse momentum of $\Xi_{bc}$. For LHCb kinematical conditions  such quark always exits in the quark sea. This is why we assume, that a doubly heavy is hadronized by joining  with a light quarks $u$, $d$ and $s$ in proportion  $1:1:0.3$. We also assume that  it is hadronized with probability equal 1. It is worth to note, that the latter assumption is  pretty much a guess, because diquark has a color charge and therefore strongly interacts with its  environment,  that could lead to the diquark dissociation. Thus, (\ref{eq:sigma_ratio}) can be considered as an upper limit for ratio of yields of $\Xi_{bc}$ and $B_c$.

\begin{figure}
  \centering
 \includegraphics[width=0.6\textwidth]{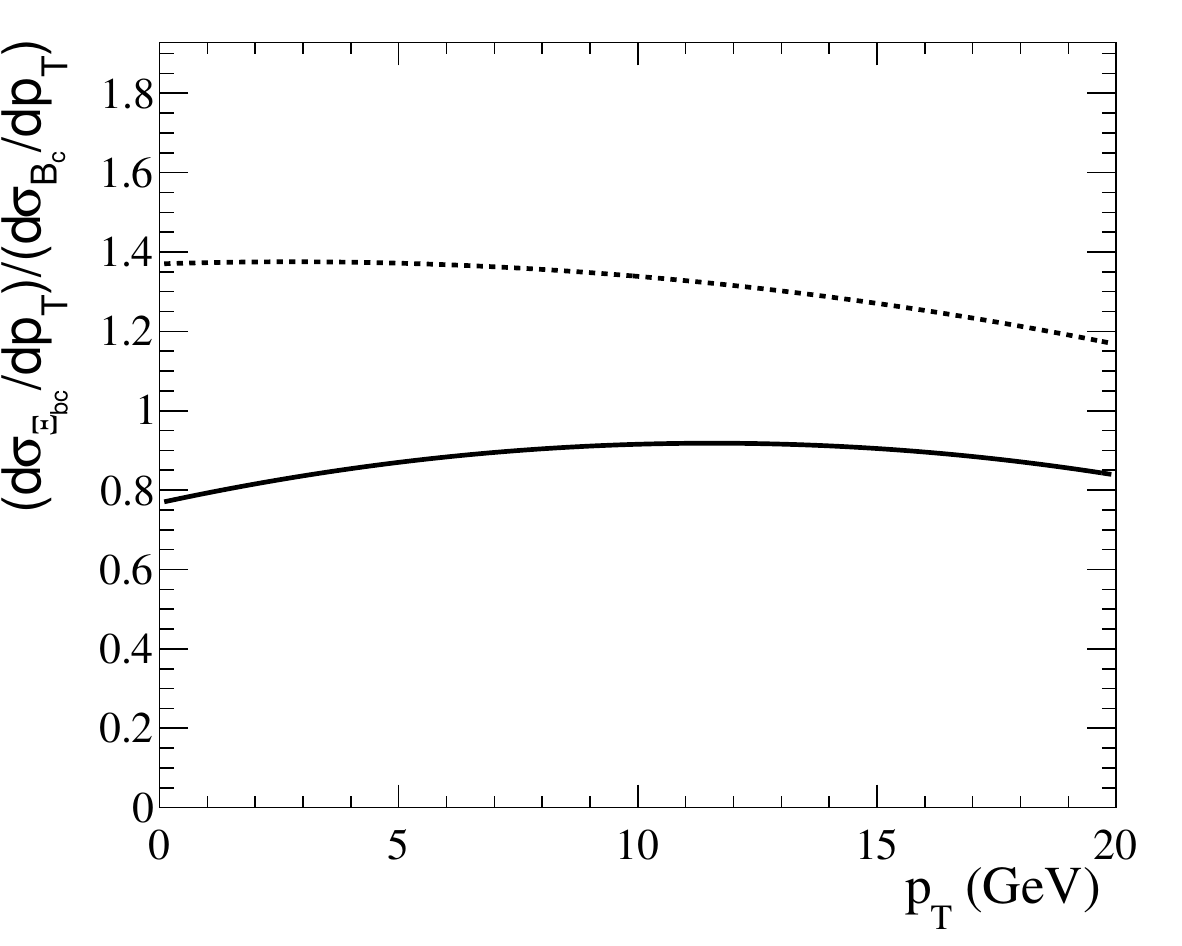}
  \caption{The ratio of production yields of $\Xi_{bc}$ and $B_c$ for hadronic interaction at $\sqrt{s}=13$~TeV  in units of $|R_{[bc]_{\bar 3}}(0)|^2/|R_{B_c}(0)|^2$ for the  similar  quark masses ($m_b=4.8$~GeV, $m_c=1.5$~GeV, solid curve) and for the different quark masses ($m_b=4.8$~GeV and $m_c=1.5$~GeV for $B_c$ production, and $m_b=4.9$~GeV and $m_c=1.7$~GeV for $\Xi_{bc}$ production, dashed curve). The CT14LL parameterization \cite{Dulat:2015mca} is used for PDFs.}
  \label{fig:ptRatio}
\end{figure}

We estimate  the ratio of yields $\Xi_{bc}$ and $B_c$  for 
hadronic interactions  at $\sqrt{s}=13$~TeV for several 
scales  ($\mu_R=\mu_F=10$~GeV, $\mu_R=\mu_F=E_T^{\Xi_{bc}}/2$, $\mu_R=\mu_F=E_T^{\Xi_{bc}}$,  $\mu_R=\mu_F=2E_T^{\Xi_{bc}}$) and find, that the dependence of this value on scale choice is unessential. The main uncertainties come from wave functions and from choice of mass values for $b$ and $c$ quarks.
In Fig.~\ref{fig:ptRatio} we show the ratio of yields $\Xi_{bc}$ and $B_c$ in hadronic interactions as a function of $p_T$ at $\sqrt{s}=13$~TeV, for the similar  masses ($m_b=4.8$~GeV, $m_c=1.5$~GeV) and for different masses ($m_b=4.8$~GeV and $m_c=1.5$~GeV for $B_c$ production, and $m_b=4.9$~GeV and $m_c=1.7$~GeV for $\Xi_{bc}$ production).  Here we also put  $|R_{B_c}(0)|^2=|R_{[bc]_{\bar 3}}(0)|^2$. One can see, that 
these distributions are  approximately flat. Thus, one can conclude, that the estimation~(\ref{eq:sigma_ratio}) is approximately valid for all transverse momenta. 


There are many estimations for  $R_{[bc]_3}(0)$ value, as well as for  $R_{B_c}(0)$ (see, for example  \cite{Gershtein:1997qy,Kiselev:2001fw,Ebert:2002ig,Ebert:2002pp,Godfrey:2004ya}). However, to obtain  the ratio, it is rational to use values  extracted within the similar framework.  From \cite{Kiselev:2001fw} and \cite{Gershtein:1997qy}, where the non-relativistic  model with Buchm\"uller-Tye wave function was used, we obtain that

$$ \frac{|R_{[bc]_{\bar 3}}(0)|^2}{|R_{B_c}(0)|^2}=
\frac{(0.71 \mbox{ GeV}^{3/2})^2}{(1.28  \mbox{ GeV}^{3/2})^2}\approx0.31.$$

From \cite{Ebert:2002pp} and \cite{Ebert:2002ig}, where the relativistic  potential model   was applied and relativistic correction have been accounted perturbatively, we obtain for the same ratio 
$$ \frac{|R_{[bc]_{\bar 3}}(0)|^2}{|R_{B_c}(0)|^2}=
\frac{(0.74 \mbox{ GeV}^{3/2})^2}{(1.46 \mbox{ GeV}^{3/2})^2}\approx 0.26.$$

In \cite{Ebert:2011jc,Galkin:2018pc} the corrections to the relativistic  potential model  predictions had been taken into account non-perturbatively, that leads to the 
noticeable difference of wave function values for different spin states. However the cross section ratio  value remains the same:   
$$
\frac{\sigma_{\Xi_{bc}}}{\sigma_{B_c}} =\frac{\sigma_{\Xi_{bc}(1^1S_0)}+\sigma_{\Xi_{bc}(1^3S_1)}}{\sigma_{B_c(1^1S_0)}+\sigma_{B_c(1^3S_1)}}\approx \frac{|R_{[bc]_{\bar 3}(1^1S_0)}(0)|^2+3\cdot|R_{[bc]_{\bar 3}(1^3S_1)}(0)|^2}{|R_{B_c(1^1S_0)}(0)|^2+2.5\cdot|R_{B_c(1^3S_1)}(0)|^2 }
\approx $$
$$\approx\frac{(0.84 \mbox{ GeV}^{3/2})^2+3\cdot( 0.59\mbox{ GeV}^{3/2})^2}{(1.64 \mbox{ GeV}^{3/2})^2+2.5\cdot(1.05\mbox{ GeV}^{3/2})^2}\approx 0.32$$

Therefore, one can conclude that
\begin{equation}
\frac{\sigma_{\Xi_{bc}}}{\sigma_{B_c}}\lesssim \frac{1}{3}.
\label{eq:sigma_ratio_numerical}
\end{equation}

It is worth to note that both the numerator and the denominator in  (\ref{eq:sigma_ratio_numerical}) will be modified by the feed-down from excitations. However we believe,  that in ratio these contributions will  approximately canceled out. The obtained ratio value $\sigma_{\Xi_{bc}}/\sigma_{B_c}$ coincides with  that used in  talk~\cite{Blusk:HHS2017}.

To estimate the absolute cross section value of $\Xi_{bc}$ baryon production at LHCb ($\sqrt{s}=13$~TeV, $2.0<y_{\Xi_{bc}}<4.5$) we use  the quark mass values $m_b=4.9$~GeV and $m_c=1.7$~GeV, the value of diquark wave function at origin $R_{[bc]_{\bar 3}}(0)= 0.71 \mbox{ GeV}^{3/2}$~\cite{Kiselev:2001fw} and CT14LL parton density parameterization \cite{Dulat:2015mca}. Varying scales from $\mu_R=\mu_F=E_T^{\Xi_{bc}}/2$ to   $\mu_R=\mu_F=2E_T^{\Xi_{bc}}$ we obtain, that  the cross section value of $bc$ baryons with $1S$ wave state of doubly heavy diquark at LHCb is about $10\div 25$ nb depending on scale values. The feed-down from excitations can be estimated as 20-30 \%.

As it was mentioned before an analogous ratio can  not be valid for $J/\psi+c$ and $\Xi_{cc}$ due to the   large contribution of DPS to the associative $J/\psi$ and $c$ production. 

\section{Doubly heavy baryon decays within OPE method}
\label{sec:lifetimes}

\subsection{Method description}
\label{sec:method}

In accordance with Operator Product Expansion (OPE) and optic theorem  the life time of doubly heavy baryon $\B$  can be represented as 
\begin{align}
  \label{eq:optical}
  \Gamma_\B &= \frac{1}{2M_\B} \left<\B \left|\T\right|\B\right>,
\end{align}
where  operator $\T$ is
\begin{align}
  \label{eq:T}
\T &= \Im\int d^4x\left\{\hat{T} H_\eff(x)H_\eff(0)\right\},
\end{align}
with
\begin{align}
  \label{eq:Heff}
  H_\eff &= \frac{G_F}{2\sqrt{2}}V_{q_3q_4}V^*_{q_1q_2}\left[C_{+}(\mu)O_{+}+C_{-}(\mu) O_{-}\right],
\end{align}
In the above expression Wilson coefficients $C_\pm(\mu)$ equal
\begin{align}
  \label{eq:C}
  C_{+}(\mu) &= \left[\frac{\alpha_s(M_W)}{\alpha_s(\mu)}\right]^{\frac{6}{33-2n_f}},
  \qquad
  C_{-}(\mu) = \left[\frac{\alpha_s(M_W)}{\alpha_s(\mu)}\right]^{-\frac{12}{33-2n_f}},
\end{align}
where $\alpha_s(\mu)$ is a running strong coupling constant calculated within two-loop approximation and $n_f$ is a number of active flavors.  The operators  $O_\pm$ in (\ref{eq:Heff}) are determined as follows:
\begin{align}
  \label{eq:O}
  O_{\pm} &= 
            \left[ \bar{q}_{1\alpha}\gamma_\nu(1-\gamma_5)q_{2\beta}\right]
            \left[ \bar{q}_{3\gamma} \gamma^\nu(1-\gamma_5) q_{4\delta}\right]
            \left( \delta_{\alpha\beta}\delta_{\gamma\delta}\pm\delta_{\alpha\delta}\delta_{\beta\gamma}\right),
\end{align}
where $\alpha$, $\beta$, $\gamma$, $\delta$ are color indices of quarks.

For large energy of heavy quark decay one  can represent    $\T$  (\ref{eq:T}) a set of local operators ordered by increasing of their dimension.
The contribution of high dimension term are suppressed by inverse powers of heavy quark mass    $m_Q$, and therefore only several first terms contribute to the decay value. This method was broadly used for the calculation of lifetimes of heavy hadrons \cite{Shifman:1984wx,Shifman:1986mx,Voloshin:1996vb,Guberina:1986gd,Falk:1993gb,
Berezhnoy:1995fy,Kiselev:1994pu,Berezhnoy:1998aa},
as well as doubly heavy hadrons \cite{Kiselev:1998sy,Kiselev:1999kh}. It was shown in  the cited papers the operators of dimension  3 and 5
\begin{align}
  \label{eq:dim35}
  O_{QQ} &= (\bar{Q} Q),\qquad O_{QG}= (\bar{Q} \sigma_{\mu\nu}G^{\mu\nu}Q), 
\end{align}
correspond to  the spectator decay of heavy quark and give the main contribution to the value (\ref{eq:optical}). The following operator of dimension 6 can also give noticable contribution to the decay process:
\begin{align}
  \label{eq:dim6}
  O_{2Q2q} &= (\bar{Q}\Gamma q)(\bar{q}\gamma Q).
\end{align}
The other operators of dimension: $O_{61Q}=\bar{Q}\sigma_{\mu\nu}\gamma_\lambda D^\mu G^{\nu\lambda} Q$, $O_{62Q}=\bar{Q}D_\mu G^{\mu\nu}\Gamma_\nu Q$,
contribute insignificantly comparing with (\ref{eq:dim6}).

\begin{figure}
  \centering
  \includegraphics[width=0.7\textwidth]{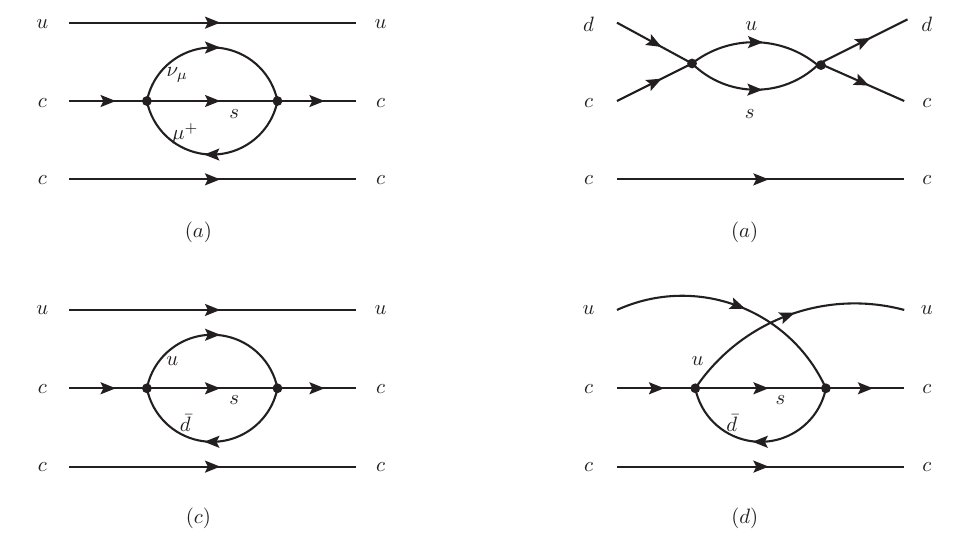}
 \caption{Feynman diagrams for $\Xi_{cc}$ baryons decay: spectator mechanism (a), weak scattering (b) and Pauly interference (c,d).}
  \label{fig:diags}
\end{figure}

Typical Feynman diagrams for the discussed processes are shown in Fig.~\ref{fig:diags}. In accordance with OPE method the following mechanisms can contribute to the total decay width: 
\begin{itemize}
\item Spectator mechanism ( the operator (\ref{eq:dim35}) and the diagram \ref{fig:diags}(a)),
\item Weak scattering,WS (the operator (\ref{eq:dim6}) and the diagram \ref{fig:diags}(b)),
\item Pauli-interference, PI (the operator (\ref{eq:dim6}) and the diagrams \ref{fig:diags}(c),~(d)),
\end{itemize}


\subsection{Lifetimes of doubly charmed baryons  $\Xi_{cc}^{++}$, $\Xi_{cc}^{+}$, $\Omega_{cc}^{+}$}
\label{sec:cc}

The decay amplitudes for  doubly charmed baryons  $\Xi_{cc}^{++}$ and $\Xi_{cc}^{+}$ can be performed as follows:
\begin{align*}
  \T_{\Xi_{cc}^{++}} &= 2 \T_{35c} + \T_{\PI}^{(\Xi_{cc}^{++})},\\
  \T_{\Xi_{cc}^{+}} &= 2 \T_{35c} + \T_{\WS}^{(\Xi_{cc}^{+})},\\
  \T_{\Omega_{cc}^{+}} &= 2 \T_{35c} + \T_{\PI}^{(\Omega_{cc}^{+})}.
\end{align*}

In these equations the contribution of operators with dimension 3 and 5 can be determined as follows:
\begin{align}
  \label{eq:T35}
  \T_{35c} &= \Gamma_{c,spec} (\bar{c}c) - \frac{\Gamma_{0c}}{m_c^2}
             \left[ (2+K_{0c})P_{s1}+K_{2c}P_{s2} \right] O_{Gc},
\end{align}
where
\begin{align*}
  \Gamma_{0c} &= \frac{G_F^2m_c^5}{192{\pi}^3},\qquad
  K_{0Q} = C_{-}^2 + 2C_{+}^2,\qquad K_{2Q} = 2(C_{+}^2 - C_{-}^2)\\
  P_{c1} &= (1-y)^4,\qquad P_{c2} = (1-y)^3, \qquad y=\frac{m_s^2}{m_c^2},\qquad r=\frac{m_\tau^2}{m_c^2}\\
  P_{c\tau 1} &= \sqrt{1-2(r+y)+(r-y)^2}[1 - 3(r+y) + 3(r^2+y^2) - r^3 - y^3 -
                \nonumber\\
         & -4ry +7ry(r+y)] + 12r^2y^2\ln\frac{(1-r-y+\sqrt{1-2(r+y)+(r-y)^2})^2}{4ry},\\
  P_{cc1} &= \sqrt{1-4y}(1 - 6y + 2y^2 + 12y^3) 24y^4\ln\frac{1+\sqrt{1-4y}}{1-\sqrt{1-4y}}\\
  P_{cc2} &= \sqrt{1-4y}(1 + \frac{y}{2} + 3y^2)- 3y(1-2y^2)\ln\frac{1+\sqrt{1-4y}}{1-\sqrt{1-4y}},
\end{align*}
and the width of spectator mechanism was estimated in papers \cite{Bigi:1992su,Bigi:1993fm,Blok:1992hw,Manohar:1993qn,Falk:1994gw,Koyrakh:1993pq, Kiselev:1998sy}.

As it was mention above the contribution values  
of $\PI$ and $\WS$ mechanisms depend on the baryon composition.  For example, it is clear from diagrams in Fig. \ref{fig:diags} that for $\Xi_{cc}^{++}=(ccu)$ and $\Omega_{cc}^{+}=(ccs)$ the WS  is forbidden  and PI destructively contributes to the width. Contrary, for the  $\Xi_{cc}^{+}$ the PI is forbidden. Taking this in mind one can perform  the contributions of  operators of 6 dimension as follows:
\begin{align*}
\T_{\PI}^{(\Xi_{cc}^{++})} &= 2 \T_{\PI,u\bar d}^c\\
\T_{\WS}^{(\Xi_{cc}^{+})} &= 2 \T_{\WS,cd}\\
\T_{\PI}^{(\Omega_{cc}^{+})} &= 2 \T_{\PI,u\bar d}^{c'} + 2 \sum_{l} \T_{\PI,\nu_l \bar l}^c  
\end{align*}
where (see, e.g,  \cite{Kiselev:1998sy,Onishchenko:1999yu,Guberina:1999mx,Guberina:1999qp})
\begin{align}
    \T_{\PI,u\bar d}^c&=  -\frac{G_F^2}{4\pi}m_c^2\left(1-\frac{m_u}{m_c}\right)^2
  \nonumber\\ &\left\{
       \left[G_{1}(z_{-})(\bar{c}c)^{ii}_{V-A}(\bar{u}u)^{jj}_{V-A}+G_{2}(z_{-})(\bar{c}c)^{ii}_{A}(\bar{u}u)^{jj}_{V-A}\right]
       \left[F_{3}+\frac{1}{3}(1-k^{\frac{1}{2}})F_{4}\right]+
       \right.
  \nonumber \\ &\left.
                 \left[G_{1}(z_{-})(\bar{c}c)^{ij}_{V-A}(\bar{u}u)^{ji}_{V-A}+G_{2}(z_{-})(\bar{c}c)^{ij}_{A}(\bar{u}u)^{ji}_{V-A}\right]k^{\frac{1}{2}}F_{4}\right\},
                 \label{a8}
  \\
  \T_{\WS,cd}&=\frac{G_F^2}{4\pi}m_c^2(1+\frac{m_d}{m_c})^2(1-z_{+})^2
               [(F_{6}+\frac{1}{3}(1-k^{\frac{1}{2}})F_{5})(\bar{c}c)^{ii}_{V-A}(\bar{d}d)^{jj}_{V-A}+
               \nonumber\\ &
               k^{\frac{1}{2}}F_{5}(\bar{c}c)^{ij}_{V-A}(\bar{d}d)^{ji}_{V-A}],
                \label{a13}\\
   \T_{\PI,u\bar d}^{c'} &=-\frac{G_F^2}{4\pi}m_c^2\left(1-\frac{m_s}{m_c}\right)^2
        \nonumber\\ &\left\{
           \left[\frac{1}{4}(\bar{c}c)^{ii}_{V-A}(\bar{s}s)^{jj}_{V-A}+\frac{1}{6}(\bar{c}c)^{ii}_{A}(\bar{s}s)^{jj}_{V-A}\right]
                      \left[F_{1}+\frac{1}{3}(1-k^{\frac{1}{2}})F_{2}\right]+
                      \right.\nonumber \\ &\left.
                      \left [\frac{1}{4}(\bar{c}c)^{ij}_{V-A}(\bar{s}s)^{ji}_{V-A}+\frac{1}{6}(\bar{c}c)^{ij}_{A}(\bar{s}s)^{ji}_{V-A}\right]k^{\frac{1}{2}}F_{2}\right\},
  \label{a9}
  \\
  \T_{\PI,\nu_{\tau}\bar\tau}^c&=-\frac{G_F^2}{\pi}m_c^2(1-\frac{m_s}{m_c})^2
                                \left[G_{1}(z_{\tau})(\bar{c}c)^{ij}_{V-A}(\bar{s}s)^{ji}_{V-A}+G_{2}(z_{\tau})(\bar{c}c)^{ij}_{A}(\bar{s}s)^{ji}_{V-A}\right],
                                   \label{a10}\\
 \T_{\PI,\nu_e\bar e}^c &= \T_{\PI,\nu_{\mu}\bar\mu}^c = \T_{\PI,\nu_{\tau}\bar\tau}^c~(z_{\tau}\to 0)\nonumber
\end{align}
and  
\begin{align}
  (\ref{a8}):\qquad &  z_{-} = \frac{m_s^2}{(m_c-m_u)^2},\quad k =\frac{\alpha_s(\mu)}{\alpha_s(m_c-m_u)},\nonumber\\
  (\ref{a13}) :\qquad &  z_{+} = \frac{m_s^2}{(m_c+m_d)^2},\quad k =\frac{\alpha_s(\mu)}{\alpha_s(m_c+m_d)}.\nonumber\\
  (\ref{a9}) :\qquad &   k =\frac{\alpha_s(\mu)}{\alpha_s(m_c-m_s)}.\nonumber\\
  (\ref{a10}) :\qquad &  z_{\tau} =\frac{m_{\tau}^2}{(m_c-m_s)^2},\nonumber
\end{align}
In these relations we also introduce the notations
\begin{align*}
  F_{1,3} &= (C_{+}\mp C_{-})^2,
            \qquad F_{2,4}=5C_{+}^2+C_{-}^2\pm 6C_{+}C_{-},
            \quad F_{5,6}=C_{+}^2\mp C_{-}^2\\
  G_1(z) &= \frac{(1-z)^2}{2}-\frac{(1-z)^3}{4},
           \quad G_2(z) = \frac{(1-z)^2}{2}-\frac{(1-z)^3}{3},\\
  (\bar{q}q)^{ij}_{A} &= (\bar{q}^i \gamma_\alpha\gamma_5 q^j),\qquad   (\bar{q}q)^{ij}_{V-A} = (\bar{q}^i \gamma_\alpha(1-\gamma_5) q^j).
\end{align*}
The hadronic matrix elements are determined as follows: 
\begin{align*}
  \langle \Xi_{QQ}^{\diamond}|(\bar Q\gamma_{\mu}(1-\gamma_5)Q)(\bar q\gamma^{\mu}(1-\gamma_5)q)|\Xi_{QQ}^{\diamond}\rangle  &=
                                                                                 12(m_Q+m_q)\cdot|\Psi^{dl}(0)|^2,\\
  \langle \Xi_{QQ}^{\diamond}|(\bar Q\gamma_{\mu}\gamma_5 Q)(\bar q\gamma^{\mu}(1-\gamma_5)q)|\Xi_{QQ}^{\diamond}\rangle
                                       &= 8(m_Q+m_q)\cdot |\Psi^{dl}(0)|^2,\\
\langle \Omega_{QQ}|(\bar Q\gamma_{\mu}(1-\gamma_5)Q)(\bar
s\gamma^{\mu}(1-\gamma_5)s)|\Omega_{QQ}\rangle  &= 12(m_Q+m_s)\cdot
|\Psi^{dl}(0)|^2,\\
\langle \Omega_{QQ}|(\bar Q\gamma_{\mu}\gamma_5 Q)(\bar
s\gamma^{\mu}(1-\gamma_5)s)|\Omega_{QQ}\rangle  &= 8(m_Q+m_s)\cdot
                                                  |\Psi^{dl}(0)|^2,
\end{align*}
where $Q=c,b$ is a heavy quark, $q=u,d$ is light quark, and $|\Psi^{dl}(0)|^{2}$ is a wave function at origin.   The wave function structure leads to the following relation: 
$$\langle \Xi_{QQ'}^{\diamond}|(\bar Q_iT_{\mu}Q_k)(\bar
q_k\gamma^{\mu}(1-\gamma_5)q_i)|\Xi_{QQ'}^{\diamond}\rangle
=
-\langle \Xi_{QQ'}^{\diamond}|(\bar QT_{\mu}Q)(\bar
q\gamma^{\mu}(1-\gamma_5)q)|\Xi_{QQ'}^{\diamond}\rangle , $$
where $T_{\mu}$ is an arbitrary spinor matrix.

\subsection{Lifetimes of doubly beauty baryons $\Xi_{bb}^{0}$, $\Xi_{bb}^{-}$, $\Omega_{bb}^{-}$}
\label{sec:bb}

For the double beauty baryons $\Xi_{bb}^{0}=(bbu)$, $\Xi_{bb}^{-}=(bbd)$ and $\Omega_{bb}^{-}=(bbs)$
WS mechanism contributes only to the width of neutral states, whereas for charge states the PI mechanism contribution must be accounted for the charged states:
\begin{align*}
    \T_{\Xi_{bb}^{0}} &= 2 \T_{35b} + \T_{\WS}^{(\Xi_{bb}^{0})},\\
  \T_{\Xi_{bb}^{-}} &= 2 \T_{35b} + \T_{\PI}^{(\Xi_{bb}^{-})}, \\
  \T_{\Omega_{bb}^{-}} &= 2 \T_{35b} + \T_{\PI}^{(\Omega_{bb}^{-})}.
\end{align*}
The spectator mechanism of $b$-quark decay is described by the following operators with dimensions 3 and 5:  
\begin{align*}
    \T_{35b} &= \Gamma_{b,spec} (\bar{b}b) - \frac{\Gamma_{0b}}{m_b^2}\left[
             2P_{c1}+P_{c\tau1}+K_{0b}(P_{c1}+P_{cc1})+K_{2b}(P_{c2}+P_{cc2}
             \right] O_{Gb},
\end{align*}
where
\begin{align*}
  \Gamma_{0c} &=\frac{G_F^2m_c^5}{192{\pi}^3},     
\end{align*}
and the other functions are determined earlier. The operators of dimension 6 equal 
\begin{align*}
  \T_{\WS}^{(\Xi_{bb}^{0})} &= 2 \T_{\WS,bu},\qquad
  \T_{\PI}^{(\Xi_{bb}^{-})} = 2 \T_{\PI,d\bar u}^{b'},\qquad
\T_{\PI}^{(\Omega_{bb}^{-})} = 2 \T_{\PI,s\bar c}^{b'},
\end{align*}
where \cite{Likhoded:1999yv}
\begin{align}
  \T_{\WS,bu} &= \frac{G_F^2|V_{cb}|^2}{4\pi}m_b^2(1+\frac{m_u}{m_b})^2(1-z_{+})^2
                [(F_{6}+\frac{1}{3}(1-k^{\frac{1}{2}})F_{5})(\bar{b}b)^{ii}_{V-A}(\bar{u}u)^{jj}_{V-A}+
                \nonumber\\&
                +k^{\frac{1}{2}}F_{5}(\bar{b}b)^{ij}_{V-A}(\bar{u}u)^{ji}_{V-A}],
  \label{a12}\\
  \T_{\PI,d\bar u}^{b'}&=-\frac{G_F^2|V_{cb}|^2}{4\pi}m_b^2\left(1-\frac{m_d}{m_b}\right)^2\left\{
                         \left[G_{1}(z_{-})(\bar{b}b)^{ii}_{V-A}(\bar{d}d)^{jj}_{V-A}+G_{2}(z_{-})(\bar{b}b)^{ii}_{A}(\bar{d}d)^{jj}_{V-A}\right]
                         \right. \nonumber \\ &\left.
                        \left[F_{3}+\frac{1}{3}(1-k^{\frac{1}{2}})F_{4}\right]+
                        \left[G_{1}(z_{-})(\bar{b}b)^{ij}_{V-A}(\bar{d}d)^{ji}_{V-A}
                                                                        \right.\right.\nonumber\\&\left. \left.
                        +G_{2}(z_{-})(\bar{b}b)^{ij}_{A}(\bar{d}d)^{ji}_{V-A}\right]k^{\frac{1}{2}}F_{4}\right\},
  \label{a6}
  \\
  \T_{\PI,s\bar c}^{b'} &=
                         -\frac{G_F^2|V_{cb}|^2}{16\pi}m_b^2(1-\frac{m_s}{m_b})^2\sqrt{(1-4z_{-})}
  \nonumber\\
  &\left\{
                         \left[(1-z_{-})(\bar{b}b)^{ii}_{V-A}(\bar{s}s)^{jj}_{V-A}+\frac{2}{3}(1+2z_{-})(\bar{b}b)^{ii}_{A}(\bar{s}s)^{jj}_{V-A}\right]\left[F_{3}+\frac{1}{3}(1-k^{\frac{1}{2}})F_{4}\right]+
                       \right.  \nonumber \\ & \left.                                               \left[(1-z_{-})(\bar{b}b)^{ij}_{V-A}(\bar{s}s)^{ji}_{V-A}+\frac{2}{3}(1+2z_{-})(\bar{b}b)^{ij}_{A}(\bar{s}s)^{ji}_{V-A}\right]k^{\frac{1}{2}}F_{4}\right\},                                              
\label{a7}
\end{align}
where
\begin{align*}
  (\ref{a12}) :\qquad &  z_{+} = \frac{m_c^2}{(m_b+m_u)^2},\quad k =\frac{\alpha_s(\mu)}{\alpha_s(m_b+m_u)},  \\
  (\ref{a6}) :\qquad &  z_{-} = \frac{m_c^2}{(m_b-m_d)^2},\quad k =\frac{\alpha_s(\mu)}{\alpha_s(m_b-m_d)},\\
  (\ref{a7}) :\qquad &  z_{-} = \frac{m_c^2}{(m_b-m_s)^2},\quad k =\frac{\alpha_s(\mu)}{\alpha_s(m_b-m_s)}
\end{align*}

\subsection{ Lifetimes of $\Xi_{bc}^{+}$, $\Xi_{bc}^{0}$, $\Omega_{bc}^{0}$ baryons}
\label{sec:bc}

It can be easily seen that in the case of $\Xi_{bc}^{+}=(bcu)$, $\Xi_{bc}^{0}=(bcd)$, and $\Omega_{bc}^{0}=(bcs)$ baryons both $\PI$ and $\WS$ channels are opened. As a result, the corresponding transition amplitudes are equal to
\begin{align*}
    \T_{\Xi_{bc}^{+}} &= \T_{35b} + \T_{35c} + \T_{\PI}^{(\Xi_{bc}^{+})} + \T_{\WS}^{(\Xi_{bc}^{+})},\\
  \T_{\Xi_{bc}^{0}} &= \T_{35b} + \T_{35c} + \T_{\PI}^{(\Xi_{bc}^{0})} + \T_{\WS}^{(\Xi_{bc}^{0})}, \\
  \T_{\Omega_{bc}^{0}} &= \T_{35b} + \T_{35c} + \T_{\PI}^{(\Omega_{bc}^{0})} + \T_{\WS}^{(\Omega_{bc}^{0})},
\end{align*}
where the contributions of $c$ and $b$ quarks' spectator decays are given in the previous subsections and $\PI$, $\WS$ amplitudes are equal to
\begin{align*}
  \T_{\PI}^{(\Xi_{bc}^{+})} &= \T_{\PI,u\bar d}^c+ \T_{\PI,s\bar c}^b+\T_{\PI,d\bar u}^b + \sum_l\T_{\PI,l\bar\nu_l}^b,\\ 
\T_{\WS}^{(\Xi_{bc}^{+})} &=\T_{\WS,bu} + \T_{\WS,bc},\\ 
\T_{\PI}^{(\Xi_{bc}^{0})} &=\T_{\PI,s\bar c}^b + \T_{\PI,d\bar u}^b+\T_{\PI,d\bar u}^{b'} + \sum_l\T_{\PI,l\bar\nu_l}^b,\\ 
\T_{\WS}^{(\Xi_{bc}^{0})} &=\T_{\WS,cd} + \T_{\WS,bc},\\
\T_{\PI}^{(\Omega_{bc}^{0})} &=\T_{\PI,u\bar d}^{c'} + \sum_{l} {\cal T}_{\PI,\nu_l \bar l}^c + \T_{\PI,s\bar c}^b + \T_{\PI,d\bar u}^b +
\sum_l\T_{\PI,l\bar\nu_l}^b + \T_{\PI,s\bar c}^{b'},\\
\T_{\WS}^{(\Omega_{bc}^{0})} &=\T_{\WS,bc} +\T_{\WS,cs}.
\end{align*}
In these expressions \cite{Kiselev:1999kh}
\begin{align}
   \T_{\PI,s\bar c}^b &= -\frac{G_F^2|V_{cb}|^2}{4\pi}m_b^2\left(1-\frac{m_c}{m_b}\right)^2\left\{
                         \left[G_{1}(z_{-})(\bar{b}b)^{ii}_{V-A}(\bar{c}c)^{jj}_{V-A}+G_{2}(z_{-})(\bar{b}b)^{ii}_{A}(\bar{c}c)^{jj}_{V-A}\right]\times
  \right.\nonumber\\&\left.
                         \left[F_{1}+\frac{1}{3}(1-k^{\frac{1}{2}})F_{2}\right]+
                    \left[G_{1}(z_{-})(\bar{b}b)^{ij}_{V-A}(\bar{c}c)^{ji}_{V-A}+G_{2}(z_{-})(\bar{b}b)^{ij}_{A}(\bar{c}c)^{ji}_{V-A}\right]k^{\frac{1}{2}}F_{2}\right\},
                      \label{a4}\\
  \T_{\PI,d\bar u}^b &= \T_{\PI,s\bar c}^b~(z_{-}\to 0),\nonumber\\
    \T_{\PI,\tau\bar\nu_{\tau}}^b &=-\frac{G_F^2|V_{cb}|^2}{\pi}m_b^2\left(1-\frac{m_c}{m_b}\right)^2\left[
                                 G_{1}(z_{\tau})(\bar{b}b)^{ij}_{V-A}(\bar{c}c)^{ji}_{V-A}+G_{2}(z_{\tau})(\bar{b}b)^{ij}_{A}(\bar{c}c)^{ji}_{V-A}\right],
                                \label{a5}
  \\
    \T_{\WS,bc} &=\frac{G_F^2|V_{cb}|^2}{4\pi}m_b^2\left(1+\frac{m_c}{m_b}\right)^2(1-z_{+})^2
                  \left[(F_{6}+\frac{1}{3}(1-k^{\frac{1}{2}}) F_{5})(\bar{b}b)^{ii}_{V-A}(\bar{c}c)^{jj}_{V-A}
                  +\right.\nonumber\\&\left.
                  +k^{\frac{1}{2}}F_{5}(\bar{b}b)^{ij}_{V-A}(\bar{c}c)^{ji}_{V-A}\right],
                  \label{a11}\\
  \T_{\PI,e\bar\nu_e}^b &= \T_{\PI,\mu\bar\nu_{\mu}}^b =
\T_{\PI,\tau\bar\nu_{\tau}}^b~(z_{\tau}\to 0)\nonumber,
\end{align}
where 
\begin{align*}
  (\ref{a4}):\qquad &  z_{-} = \frac{m_c^2}{(m_b-m_c)^2},\quad k = \frac{\alpha_s(\mu)}{\alpha_s(m_b-m_c)},\\
  (\ref{a5}):\qquad &  z_{\tau} =\frac{m_{\tau}^2}{(m_b-m_c)^2},\\
  (\ref{a11}):\qquad &  z_{+} = \frac{m_c^2}{(m_b+m_c)^2},\quad k =\frac{\alpha_s(\mu)}{\alpha_s(m_b+m_c)}.
\end{align*}
The other functions are defined earlier.

\subsection{Numerical results}
\label{sec:numerical}

From presented above results it is clear that in OPE formalism theoretical predictions of doubly heavy baryons' lifetimes depend on such input parameters as quark masses, wave function at the origin, etc. In paper \cite{Likhoded:1999yv} the following values of these parameters were used:
\begin{align}
  V_{cs} &= 0.9745, \quad V_{cb}=0.04, \\
  T&=0.4\,\GeV,\qquad |\Psi^{dl}(0)|^2=(2.7\pm0.2)\times 10^{-3}\GeV^3,\\
  \label{eq:num_mass_const}
  m_s&=0.2\,\GeV,\quad m_c=1.55\,\GeV,\quad m_b=5.05\,\GeV.
\end{align}
This choice however  leads to the following values of $\Xi_{cc}^{++}$ baryon mass and lifetime:
\begin{align}
  M_{\Xi_{cc}} &= 3.478\,\GeV,\quad \tau_{\Xi_{cc}^{++}}=0.44\,\ps
\end{align}
These results, unfortunately, disagree with resent experimental data \cite{Aaij:2017ueg,Aaij:2018wzf}
\begin{align}
  \label{eq:MTexp}
  M_{\Xi_{cc}^{++}}^{\mathrm{exp}} &= (3621.40 \pm 0.72 \pm 0.27 \pm 0.14)\,\MeV,\qquad
  \tau_{\Xi_{cc}^{++}}^{\mathrm{exp}} = 0.256 ^{+0.024}_{-0.022}\pm0.014\,\ps,
\end{align}
so some change of parameters is required. It should be noted that the values \eqref{eq:num_mass_const} correspond to constituent quark masses obtained from analysis of $D$-mesons' lifetimes. In papers \cite{Karliner:2014gca,Karliner:2018hos} it was proposed that a slightly different masses should be used in the case of doubly heavy baryons. We will discuss the results of these papers in the next subsection, while here we consider quark masses as free and check the dependence of doubly heavy baryons lifetimes on the variation of these parameters.

\begin{figure}
  \centering
  \includegraphics[width=0.32\textwidth]{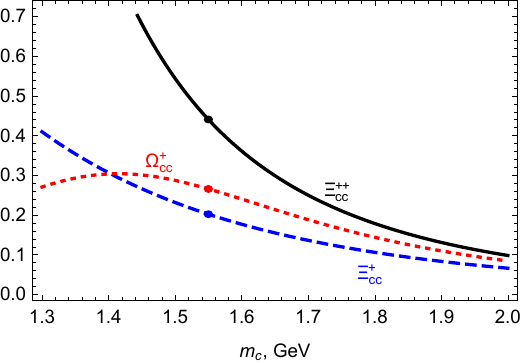}
  \includegraphics[width=0.32\textwidth]{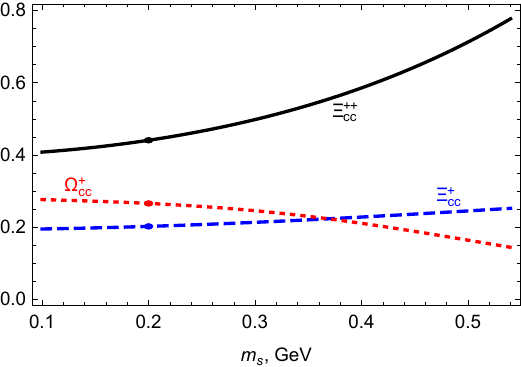}
  \includegraphics[width=0.32\textwidth]{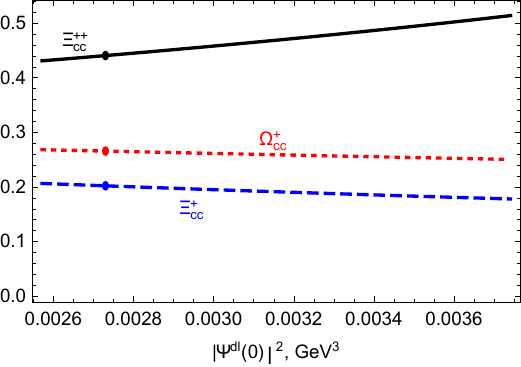}
  \caption{Lifetimes in $\ps$ for $\Xi_{cc}^{++}$ (solid black curve), $\Xi_{cc}^{+}$ (blue dashed curve)  and $\Omega_{cc}^{+}$ (red dotted curve) as a function of the model parameters.  The results of  \cite{Likhoded:1999yv} are shown by dots.}
  \label{fig:cc}
\end{figure}

\begin{figure}
  \centering
  \includegraphics[width=0.32\textwidth]{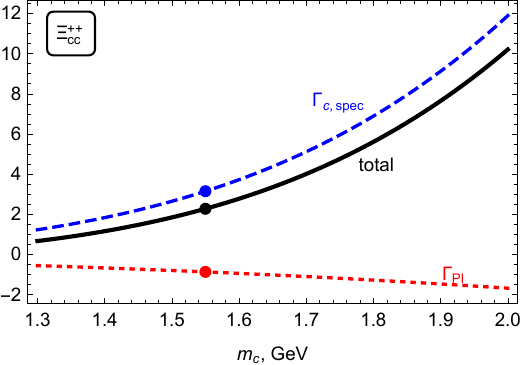}
  \includegraphics[width=0.32\textwidth]{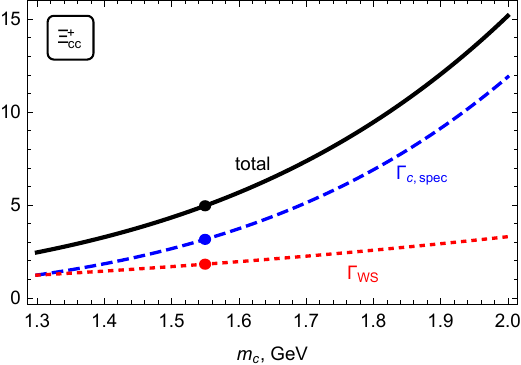}
  \includegraphics[width=0.32\textwidth]{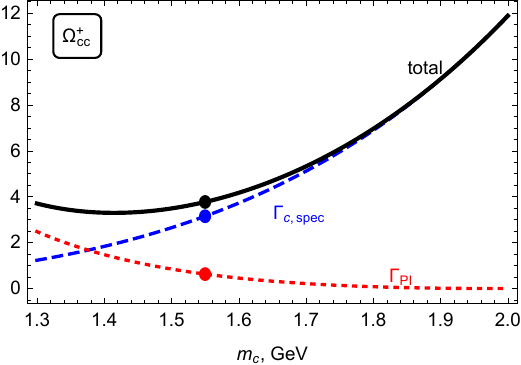}
  \caption{The partial widths for different operators for doubly charmed baryons (in $\ps^{-1}$):  operators of dimension 3 and 5  corresponds to the  spectator mechanism (dashed blue curve), operators of dimension 6 corresponding to the weak scattering and Pauly interference  (red dotted curve), the total width (black solid curve). The dots correspond to the predictions of  \cite{Likhoded:1999yv}.}
  \label{fig:cc_channels_mc}
\end{figure}

\begin{table}
  \centering
  \begin{tabular}{l|ccc}
    \hline
    & $\Xi_{cc}^{++}$ & $\Xi_{cc}^{+}$ & $\Omega_{cc}^{+}$ \\
    \hline
$\sum c\to s$, $\ps^{-1}$ & $5.1\pm0.5$ ($3.1$) & $5.1\pm0.5$ ($3.1$) & $5.1\pm0.5$ ($3.1$) \\ 
 PI, $\ps^{-1}$ & $-1.2\pm0.1$ ($-0.87$) & --- & $0.65\pm0.5$ ($0.62$) \\ 
 WS, $\ps^{-1}$ & --- & $2.3\pm0.2$ ($1.8$) & --- \\ 
\hline
 $\tau,\,\ps$ & $0.26\pm0.03$ ($0.44$) & $0.14\pm0.01$ ($0.2$) & $0.18\pm0.02$ ($0.27$) \\ 
                  \hline
    \end{tabular}
    \caption{Lifetimes of doubly charmed baryons and different and partial contributions of different mechanisms (values in brackets correspond to  \cite{Likhoded:1999yv}. Theoretical uncertainties are caused by $m_{s,c}$ variation (\ref{eq:mass_fit}).}
  \label{tab:cc_channels}
\end{table}

In Figure \ref{fig:cc} we show model parameter dependence of $\Xi_{cc}^{++}$ lifetime, while Fig.~\ref{fig:cc_channels_mc}a shows $m_{c}$ dependence of different channels that contribute to this lifetime. It can be seen from these figures that  $\tau(\Xi_{cc}^{++})$ is most sensitive to change of $c$-quark mass. Our analysis shows that experimental value~\eqref{eq:MTexp} is restored with the following values:
\begin{align}
  \label{eq:mass_fit}
  m_{c} &= 1.73\pm0.07\,\GeV,\qquad m_{s}=0.35\pm0.2\,\GeV.
\end{align}
With these masses we have $\tau(\Xi_{cc}^{++})=0.26\pm0.03\,\ps$. In the second column of table \ref{tab:cc_channels} we show calculated with these masses contributions of different decay channels to $\Xi_{cc}^{++}$ baryon lifetime in comparison with that presented in \cite{Likhoded:1999yv}. One can see from this table that, as it was mentioned in the previous sections, the spectator decay channel gives the main contribution and it increases with the increase of charm quark mass. In addition, PI channel gives destructive contribution in this case, which leads to increase of the lifetime. As for weak scattering mechanism, it is forbidden for $\Xi_{cc}^{++}$ decay.

Using  the approach described above, it is easy to calculate also lifetimes of $\Xi_{cc}^{+}$ and $\Omega_{cc}^{+}$ baryons:
\begin{align}
  \tau(\Xi_{cc}^{+}) &= 0.14\pm0.01\,\ps,\qquad \tau(\Omega_{cc}^{+})=0.18\pm0.02\,\ps.
\end{align}
Lifetime and decay width dependences on parameters are shown in figures \ref{fig:cc}, \ref{fig:cc_channels_mc}.  The numerical estimations for parameter values (\ref{eq:num_mass_const}) and (\ref{eq:mass_fit})  can be found in the third and fourth columns of table \ref{tab:cc_channels}. In the case of $\Xi_{cc}^{+}$ baryon the PI channel is forbidden, thus only the spectator decay and the weak scattering give contributions. For for $\Omega_{cc}^{+}$ baryon the spectator and PI channels are important. The contribution of the last one is positive. As a result theoretical predictions for the lifetimes of $\Xi_{cc}^{+}$ and $\Omega_{cc}^{+}$ are smaller than for $\Xi_{cc}^{++}$ particle.

\begin{figure}
  \centering
  \includegraphics[width=0.32\textwidth]{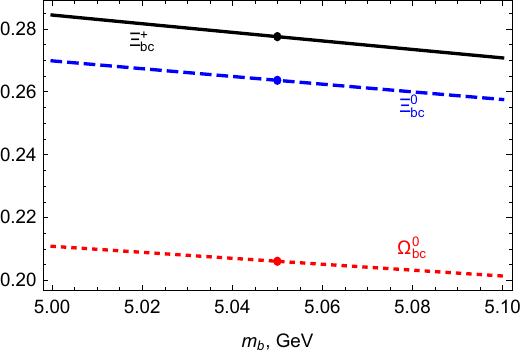}
  \includegraphics[width=0.32\textwidth]{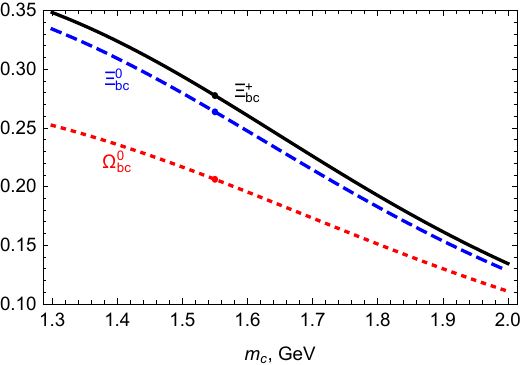}
  \includegraphics[width=0.32\textwidth]{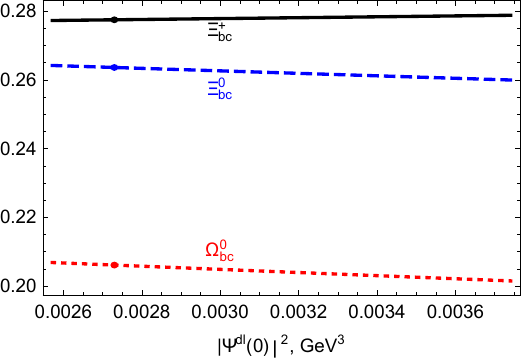}
  \caption{
 Lifetimes in $\ps$ for $\Xi_{bc}^{+}$ (solid black curve), $\Xi_{bc}^{0}$ (blue dashed curve)  and $\Omega_{bc}^{0}$ (red dotted curve) as a function of the model parameters.  The results of  \cite{Likhoded:1999yv} are shown by dots.}
  \label{fig:bc}
\end{figure}

\begin{figure}
  \centering
  \includegraphics[width=0.32\textwidth]{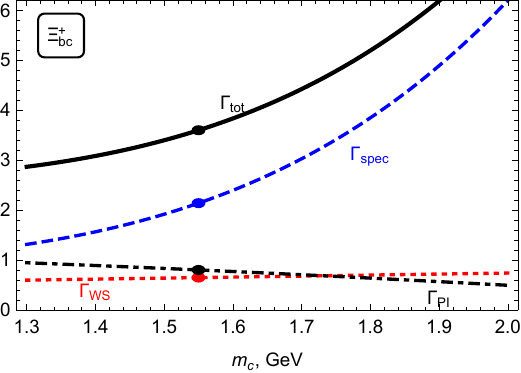}
  \includegraphics[width=0.32\textwidth]{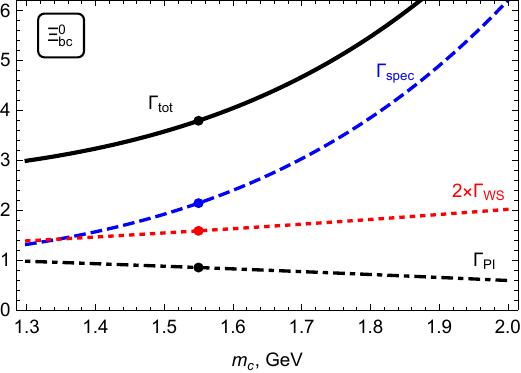}
  \includegraphics[width=0.32\textwidth]{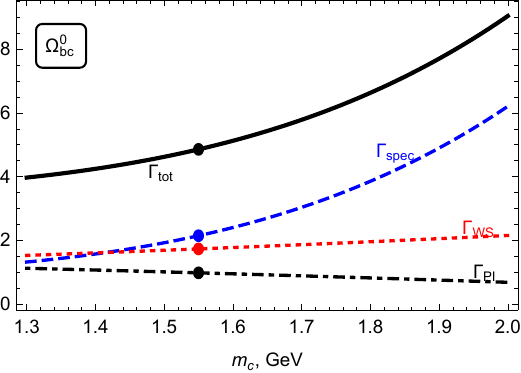}
  \caption{The partial widths for different operators for $bc$ baryons (in $\ps^{-1}$):  operators of dimension 3 and 5  corresponds to the  spectator mechanism (dashed blue curve), operators of dimension 6 corresponding to the weak scattering and Pauly interference  (red dotted and black dash-dotted curves respectively), the total width (black solid curve). The dots correspond to the predictions of  \cite{Likhoded:1999yv}.}
  \label{fig:bc_channels_mc}
\end{figure}

\begin{table}
  \centering
  \begin{tabular}{l|ccc}
    \hline
    & $\Xi_{bc}^{+}$ & $\Xi_{bc}^{0}$ & $\Omega_{bc}^{0}$ \\
    \hline
    $\sum b\to c$, $\ps^{-1}$ & $0.551\pm0.0311$ ($0.632$) & $0.551\pm0.0311$ ($0.632$) & $0.551\pm0.0311$ ($0.632$) \\ 
    $\sum c\to s$, $\ps^{-1}$ & $2.32\pm0.342$ ($1.51$) & $2.32\pm0.342$ ($1.51$) & $2.32\pm0.342$ ($1.51$) \\ 
    PI, $\ps^{-1}$ & $0.69\pm0.044$ ($0.81$) & $0.75\pm0.039$ ($0.86$) & $0.86\pm0.044$ ($0.98$) \\ 
    WS, $\ps^{-1}$ & $0.69\pm0.014$ ($0.65$) & $0.87\pm0.022$ ($0.79$) & $2.\pm0.13$ ($1.7$) \\ 
\hline
 $\tau,\,\ps$ & $0.24\pm0.02$ ($0.28$) & $0.22\pm0.018$ ($0.26$) & $0.18\pm0.0088$ ($0.21$) \\ 
    \hline
    \end{tabular}
    \caption{ Decay widths and lifetimes for $bc$-baryons. The meaning of symbols is the same as in Tab.~\ref{tab:cc_channels}}
  \label{tab:bc_channels}
\end{table}

Let us now consider lifetimes of  $bc$-baryons $\Xi_{bc}^{+}$, $\Xi_{bc}^{0}$, and $\Omega_{bc}^{0}$.  The  lifetime dependences on parameters are shown  in Figure~\ref{fig:bc}.
In the following we will use constituent value $m_{b}=5.05\,\GeV$ for $b$-quark mass and (\ref{eq:mass_fit}) for $m_{c,s}$. In Figure \ref{fig:bc_channels_mc} we show $m_{c}$ dependence of different channel contributions for these baryons. The predictions corresponding to parameter values (\ref{eq:num_mass_const}) and (\ref{eq:mass_fit})  are given in table \ref{tab:bc_channels}. From presented results it is clear, that $c$-quark spectator decay  is dominant for the considered baryons, while contributions of $b$-quark spectator decay is suppressed by $V_{cb}$ matrix element. As for dimension 6 operators PI and WS, their contributions are suppressed by large $b$-quark mass and are small. It is interesting to note, however, that, in contrast to $cc$ baryons,  in the case of $bc$-baryons both PI and WS channels are not forbidden for all considered particles.

\begin{figure}
  \centering
  \includegraphics[width=0.32\textwidth]{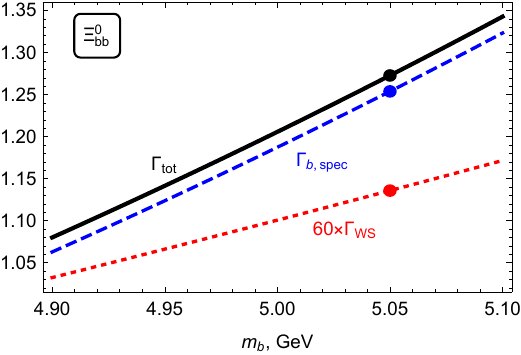}
  \includegraphics[width=0.32\textwidth]{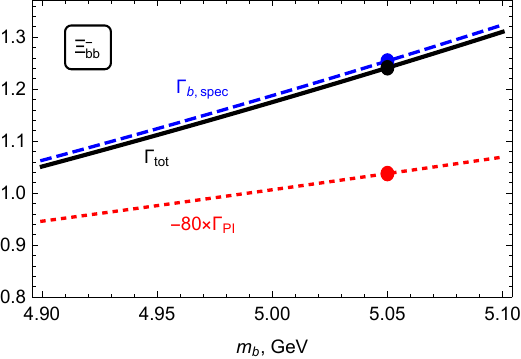}
  \includegraphics[width=0.32\textwidth]{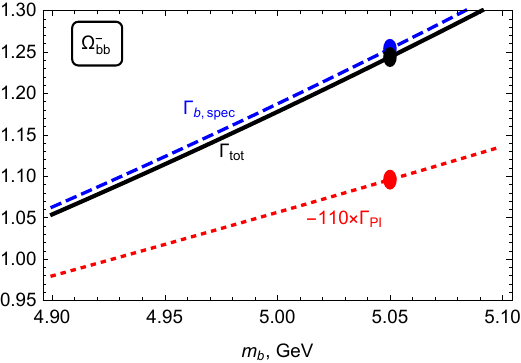}
  \caption{Decay widths for $bb$ baryons.  Designations as in Fig.~\ref{fig:cc_channels_mc}}
  \label{fig:bb_channels_mc}
\end{figure}

\begin{table}
  \centering
  \begin{tabular}{l|ccc}
    \hline
    & $\Xi_{bb}^{0}$ & $\Xi_{bb}^{-}$ & $\Omega_{bb}^{-}$ \\
    \hline
    $\sum b\to c$, $\ps^{-1}$ & $1.9\pm0.0344$ ($1.25$) & $1.9\pm0.0344$ ($1.25$) & $1.9\pm0.0344$ ($1.25$) \\ 
 PI, $\ps^{-1}$ & --- & $-0.016\pm0.0003$ ($-0.013$) & $-0.011\pm0.0014$ ($-0.01$) \\ 
 WS, $\ps^{-1}$ & $0.023\pm0.00064$ ($0.019$) & --- & --- \\ 
\hline
 $\tau,\,\ps^{-1}$ & $0.52\pm0.0095$ ($0.79$) & $0.53\pm0.0096$ ($0.81$) & $0.53\pm0.0093$ ($0.8$) \\ 

                  \hline
    \end{tabular}
    \caption{ Decay widths and lifetimes for $bb$-baryons Designations as on Tab.~\ref{tab:cc_channels}}
  \label{tab:bb_channels}
\end{table}

\begin{figure}
  \centering
  \includegraphics[width=0.32\textwidth]{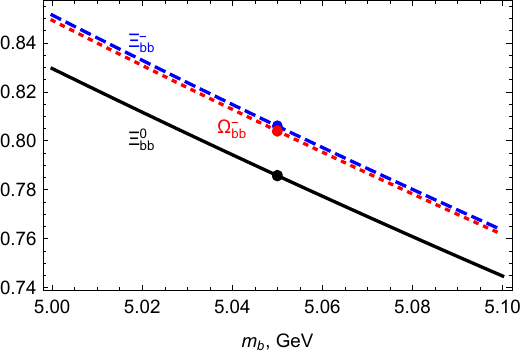}
  \includegraphics[width=0.32\textwidth]{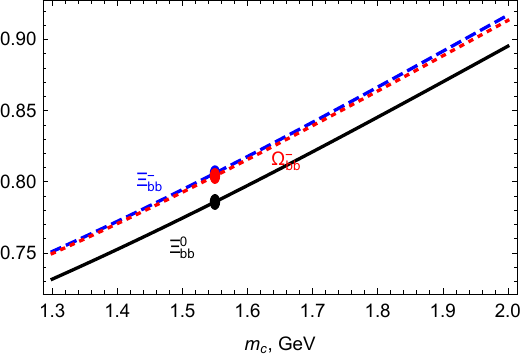}
  \includegraphics[width=0.32\textwidth]{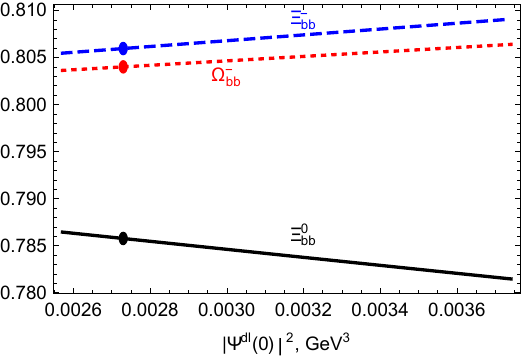}
  \caption{
 Lifetimes in $\ps$ for $\Xi_{bb}^{0}$ (solid black curve), $\Xi_{bb}^{-}$ (blue dashed curve)  and $\Omega_{bb}^{-}$ (red dotted curve) as a function of the model parameters.  The results of  \cite{Likhoded:1999yv} are shown by dots.}
  \label{fig:bb}
\end{figure}

In the case of $bb$-baryons $\Xi_{bb}^{0}$, $\Xi_{bb}^{-}$, and $\Omega_{bb}^{-}$ spectator $b$-quark decay gives the dominant contribution. As for dimension 6 operators, in complete agreement with OPE selection rules their contributions are suppressed by large quark mass. As a result, lifetime values presented in Table \ref{tab:bb_channels}  are close to each other. It should be noted that, similar to $cc$ sector, different decay mechanisms are enabled for different baryons: WS is enabled only for neutral particle and  PI is enabled only for  charged ones. Parameter dependence of the lifetimes and decay widths of these baryons are shown in figures \ref{fig:bb_channels_mc} and \ref{fig:bb}.

\subsection{Comparison with Other Works}
\label{sec:comp}

One can find in the literature some other theoretical works devoted to analysis of doubly heavy baryons lifetimes. In the current subsection we will discuss these papers and compare presented there results with ours.

As it was mentioned above, in papers \cite{Karliner:2014gca,Karliner:2018hos} it was assumed that quark masses used for doubly heavy baryons analysis could be a little bit different from constituent quark masses obtained from analysis of meson spectroscopy. In particular, in paper \cite{Karliner:2014gca} ([KR14])the following values were considered:
\begin{align}
  \label{eq:KRqmass}
  m_{q}^{\mathrm{[KR14]}} &= 363\,\MeV,\qquad m_{s}^{\mathrm{[KR14]}}=538\,\MeV,\qquad m_{c}^{\mathrm{[KR14]}}=1.7105\,\GeV, 
\end{align}
that correspond to $\Xi_{cc}^{++}$ mass and lifetime equal to
\begin{align}
  \label{eq:KRMT}
  M_{\Xi_{cc}}^{\mathrm{[KR14]}} &= (3627\pm 12)\,\MeV,\qquad
  \tau_{\Xi_{cc}^{++}}^{\mathrm{[KR14]}}=0.185\,\ps
\end{align}
One can see that the mass of the baryon is more close to the experimental value \eqref{eq:MTexp}, while the lifetime is even smaller. We would like, however, make some comments considering the last result. Presented in \cite{Karliner:2014gca} analytical expression for $\Xi_{cc}^{++}$ decay width reads
\begin{align}
  \label{eq:KRGamma}
  \Gamma_{tot}^{\mathrm{[KR14]}}(\Xi_{cc}^{++}) &=10\frac{G_{F}^{2}M_{\Xi_{cc}}^{2}}{192\pi^{3}}f(x_{cc}),\qquad
   x_{cc} = \frac{M_{\Xi_{cc}}^{2}}{M_{\Xi_{c}}^{2}}.
\end{align}
From this expression it is clear that in \cite{Karliner:2014gca} only spectator decays of the valence $c$ quark contribute. Indeed, the prefactor $10=2\times(3+1+1)$ in relation \eqref{eq:KRGamma} shows that only $c\to s u d$, $c\to s e \nu_{c}$, and $c\to\mu\nu_{mu}$ channels were taken into account and the final result is doubled because of two valence quarks in $\Xi_{cc}$ baryon Fock state. It seems to us, that such an approach is not reliable.

First of all, as it can be clearly seen from comparison with neutron's total width, mentioned above factor 2 should be avoided. Indeed, since only one spectator decay $d\to u e\nu_{e}$ is possible in this case and there are two valence $d$ quarks in the neutron, used in \cite{Karliner:2014gca} approach would give us the lifetime
\begin{align}
  \label{eq:tauN}
  \tau_{n} = \left[2\frac{G_{F}^{2}m_{n}^{5}}{192\pi^{3}}f\left(\frac{m_{p}^{2}}{m_{n}^{2}}\right)\right]^{-1}\approx 320\,\s,
\end{align}
which is almost three times smaller than the experimental result $\tau_{n}^{\mathrm{exp}}=939\,\s$. Without the factor 2 in relation \eqref{eq:tauN} this disagreement is partially removed. In addition, in paper \cite{Karliner:2014gca} contributions of any form factors are neglected. It is clear that the energy deposit in $\Xi_{cc}$ baryon decay is much larger than for neutron $\beta$-decay. It is well known, however, that even in the latter case $n\to pe\nu_{e}$ such form factors are important (actually, the axial form factor helps us to obtain the experimental value of the considered lifetime), so it seems strange to forget about them in the case of $\Xi_{cc}$ lifetime.

The other point is that PI and WS contributions are completely ignored in \cite{Karliner:2014gca}. As a result, one can expect that lifetimes of all $ccq$, $ccs$ baryons should be equal to each other. For some reason, however, the authors of paper \cite{Karliner:2014gca} use completely different approach to calculate $\Xi_{cc}^{+}$ baryon lifetime and the value $\tau_{\Xi_{cc}^{+}}\approx\tau(\Xi_{cc}^{++})/2$ is given there. No detailed explanation for such difference in calculation methods is presented in \cite{Karliner:2014gca}.

If we use the presented in [KR14] values in described above OPE calculations, the lifetime of $\Xi_{cc}^{++}$ baryon is equal to $0.32\ps$, that is a little bit larger than the experimental result \eqref{eq:MTexp}. In paper \cite{Karliner:2018hos} ([KR18]) another set of quark masses was presented, that describe both meson and baryon masses:
\begin{align}
  \label{eq:KR2qmass}
  m_{q}^{\mathrm{[KR18]}} &= 308.5\,\MeV,\qquad m_{s}^{\mathrm{[KR18]}}=482.2\,\MeV,\qquad m_{c}^{\mathrm{[KR18]}}=1655.6\,\GeV, 
\end{align}
No predictions for the lifetimes can be found in this paper, but OPE approach gives the value $\tau(\Xi_{cc}^{++})\approx0.37\,\ps$, which is also larger than the experimental one.

In a series of papers \cite{Guberina:1997yx,Guberina:1999mx,Guberina:1999bw,Guberina:2000de} the lifetimes of heavy and doubly heavy baryons are considered in the framework of operator product expansion with PI and WS channels taken into account. The result of these works agrees qualitatively with ours (for example, the hierarchy of $cc$-baryons lifetimes is the same), but the numerical values of the lifetimes are somewhat larger. The reason for the difference is that used in these papers values of quark masses are smaller (for example, $m_{c}=1.35\,\GeV$ in these papers).

It should be noted that the mass of $c$ quark is not really large, so higher order contributions in operator product expansion could also give significant contributions. In the recent article \cite{Cheng:2018mwu}  the authors show that the experimental value of $\Xi_{cc}^{++}$ baryon lifetime can be explained if contributions of higher dimension operators are taken into account. It is interesting to note, that the lifetimes of other doubly charmed baryons are changed in different way in comparison with our results: $\tau(\Xi_{cc}^{+})$ decreases only slightly, while the lifetime of $\Omega_{cc}^{+}$ baryon increases and is comparable with $\tau(\Xi_{cc}^{++})$.  It is clear that a detailed theoretical and experimental investigation of the lifetimes of these particles is highly desirable.

\section{Observation Perspectives}
\label{sec:observ-prob}

Here we briefly discuss the observation possibilities of doubly heavy baryons at LHC.
As it was already mentioned the observation of $\Xi_{cc}^{++}$ baryon has been done by  the LHCb Collaboration in the decay mode $\Lambda_c^+ K^- \pi^+ \pi^+$~\cite{Aaij:2017ueg} and  confirmed in the decay mode $\Xi_{c}^{+}\pi^{+}$~\cite{Aaij:2018gfl}.

The next step is the observation of $\Xi_{cb}$  baryon. In spite of large number of theoretical predictions for  branching fractions  (see, for example, \cite{Kiselev:2001fw,Yu:2017zst,Albertus:2012nd,Albertus:2012nc,Faessler:2001mr, Onishchenko:2000yp, Wang:2017mqp} and Table \ref{tab:exclusiveBr}), the "golden mode" is not found yet.
 Of course, the greater  branching fraction value, the more chances for the decay mode  to be observed. But the decay branchings  of intermediate particles are also very important.  In addition, as it is
shown in \cite{Blusk:HHS2017}, the possibility of the experiment also must be taken into account. For example, each extra track in final state decreases the registration efficiency. That is why understanding the experiment features  is very important for searching the most promising decay modes.   We share cautious optimism of  \cite{Blusk:HHS2017} about the observation of particle in the LHCb data of Run I and Run II, and also think that 	
in any case $\Xi_{cb}$  will be observed in the LHCb data of Run III. 

As for the observation of the $\Xi_{bb}$, we doubt its possibility at the LHC because of the very small  production rate. 

\begin{table}
  \centering
  \begin{tabular}{c|c|c||c|c|c||c|c|c}
    Mode & \cite{Onishchenko:2000yp, Kiselev:2001fw} &  \cite{Wang:2017mqp} &
                                                              Mode & \cite{Onishchenko:2000yp,Kiselev:2001fw} & \cite{Wang:2017mqp} &
                                                                                                                       Mode & \cite{Onishchenko:2000yp,Kiselev:2001fw} & \cite{Wang:2017mqp} \\
    \hline
    $\Xi_{cc}^{++}\to \Xi_{c}^{+} \rho^{+}$ & 46.8 & 14.2 & $\Xi_{cc}^{+}\to \Xi_{c}^{0} \rho^{+}$ & 33.6 & 4.66 & $\Omega_{cc}^{+}\to \Omega_{c}^{0} \rho^{+}$ & --- & 24.2 \\ 
    $\Xi_{cc}^{++}\to \Xi_{c}^{+} \pi$ & 15.7 & 7.24 & $\Xi_{cc}^{+}\to \Xi_{c}^{0} \pi$ & 11.2 & 2.4 & $\Omega_{cc}^{+}\to \Omega_{c}^{0} \pi$ & --- & 7.05 \\ 
    $\Xi_{cc}^{++}\to \Xi_{c}^{+} \ell\nu_{\ell}$ & 16.8 & 5.39 & $\Xi_{cc}^{+}\to \Xi_{c}^{0} \ell\nu_{\ell}$ & 7.5 & 1.77 & $\Omega_{cc}^{+}\to \Omega_{c}^{0} \ell\nu_{\ell}$ & --- & 6.65 \\ 
    \hline
    $\Xi_{bc}^{+}\to \Xi_{b}^{0} \rho^{+}$ & 21.7 & 6.24 & $\Xi_{bc}^{0}\to \Xi_{b}^{-} \rho^{+}$ & 20.1 & 2.36 & $\Omega_{bc}^{0}\to \Omega_{b}^{-} \rho^{+}$ & --- & 18. \\ 
    $\Xi_{bc}^{+}\to \Xi_{b}^{0} \pi$ & 7.7 & 3.25 & $\Xi_{bc}^{0}\to \Xi_{b}^{-} \pi$ & 7.1 & 1.23 & $\Omega_{bc}^{0}\to \Omega_{b}^{-} \pi$ & --- & 4.57 \\ 
    $\Xi_{bc}^{+}\to \Xi_{b}^{0} \ell\nu_{\ell}$ & 4.4 & 2.3 & $\Xi_{bc}^{0}\to \Xi_{b}^{-} \ell\nu_{\ell}$ & 4.1 & 0.867 & $\Omega_{bc}^{0}\to \Omega_{b}^{-} \ell\nu_{\ell}$ & --- & 6. \\ 
    \hline
    $\Xi_{bb}^{0}\to \Xi_{bc}^{+} \ell\nu_{\ell}$ & 14.9 & 2.59 & $\Xi_{bb}^{-}\to \Xi_{bc}^{0} \ell\nu_{\ell}$ & 14.9 & 1.68 & $\Omega_{bb}^{-}\to \Omega_{bc}^{0} \ell\nu_{\ell}$ & --- & 4.83 \\ 
    $\Xi_{bb}^{0}\to \Xi_{bc}^{+} \rho^{-}$ & 5.7 & 0.617 & $\Xi_{bb}^{-}\to \Xi_{bc}^{0} \rho^{-}$ & 5.7 & 0.265 & $\Omega_{bb}^{-}\to \Omega_{bc}^{0} \rho^{-}$ & --- & 1.25 \\ 
    $\Xi_{bb}^{0}\to \Xi_{bc}^{+} \pi$ & 2.2 & 0.213 & $\Xi_{bb}^{-}\to \Xi_{bc}^{0} \pi$ & 2.2 & 0.0854 & $\Omega_{bb}^{-}\to \Omega_{bc}^{0} \pi$ & --- & 0.43 \\ 
    \hline
  \end{tabular}
  \caption{Branching fractions of the exclusive decays}
  \label{tab:exclusiveBr}
\end{table}

\section{Conclusions}
\label{sec:conclusion}

This article is devoted to theoretical study of total widths, production rates, and observation probabilities of the doubly heavy baryons.

We briefly discussed the production and the possibility of observation  of $\Xi_{bc}$ baryon at LHC, and showed that the kinematical features  of $\Xi_{bc}$ baryon production and $B_c$ meson production are very similar.

The main efforts were made to estimate the lifetimes of doubly heavy baryons in the framework of Operator Product Expansion (OPE). 
We studied the lifetime dependence on main parameters of this formalism, which are masses of $s$, $c$, and $b$ quarks and the value of the diquark wave function at the origin.
We show, that the spectator heavy quark decays give the main contribution to the lifetimes of doubly heavy baryons. However, in the case of $\Xi_{cc}$ and $\Omega_{cc}$ baryons the contributions of the higher dimension terms, such as weak scattering and Pauli interference channels, are also important.  For  $bcq$ and $bbq$ baryons the higher dimension terms are suppressed by the large mass of the heavy quark and do not contribute essentially to the lifetime value. 

The lifetime predictions for doubly heavy baryons are most sensitive to  the charm quark mass. The knowledge of the experimental value of $\Xi_{cc}^{++}$ baryon lifetime  allowed us to determine this parameter with  pretty good accuracy and to make the lifetime predictions for other doubly heavy baryons.

The authors would like to thank V. Galkin, V.V. Kiselev,  and A. Onishchenko for help and useful discussions. The work was carried out with the financial support of  RFFBR (grant 19-02-00302).
A.Berezhnoy  also acknowledges the support from MinES of RF (grant 14.610.21.0002, identification number RFMEFI61014X0002), "Basis" Foundation (grant 17-12-244-1).

\bibliographystyle{apsrev4-1}
\bibliography{dhb-litr}

\end{document}